# Dark matter maps reveal cosmic scaffolding

Richard Massey[1], Jason Rhodes[2,1], Richard Ellis[1], Nick Scoville[1], Alexie Leathaud[3], Alexis Finoguenov[4], Peter Capak[1], David Bacon[5], Hervé Aussel[6], Jean-Paul Kneib[3], Anton Koekemoer[7], Henry McCracken[8], Bahram Mobasher[7], Sandrine Pires[6], Alexandre Refregier[6], Shunji Sasaki[9], Jean-Luc Starck[6], Yoshi Taniguchi[9], Andy Taylor[5] & James Taylor[10*]

**Ordinary baryonic particles (such as protons and neutrons) account for only one-sixth of the total matter in the Universe[1–3]. The remainder is a mysterious 'dark matter' component, which does not interact via electromagnetism and thus neither emits nor reflects light. As dark matter cannot be seen directly using traditional observations, very little is currently known about its properties. It does interact via gravity, and is most effectively probed through gravitational lensing: the deflection of light from distant galaxies by the gravitational attraction of foreground mass concentrations[4,5]. This is a purely geometrical effect that is free of astrophysical assumptions and sensitive to all matter—whether baryonic or dark[6,7]. Here we show high-fidelity maps of the large-scale distribution of dark matter, resolved in both angle and depth. We find a loose network of filaments, growing over time, which intersect in massive structures at the locations of clusters of galaxies. Our results are consistent with predictions of gravitationally induced structure formation[8,9], in which the initial, smooth distribution of dark matter collapses into filaments then into clusters, forming a gravitational scaffold into which gas can accumulate, and stars can be built[10].**

The *Hubble Space Telescope* (HST) *Cosmic Evolution Survey* (COSMOS) is the largest contiguous expanse of high-resolution imaging data obtained from space[11]. 575 slightly overlapping pointings of the *Advanced Camera for Surveys* (ACS) *Wide Field Camera* cover a region of 1.637 square degrees. We measure the shapes of half a million distant galaxies[12], and use their observed distortion (*c.f.* supplementary Fig 1.) to reconstruct the distribution of intervening mass, projected along our line of sight (Fig.1). A realisation of noise in our mass map, including most spurious instrumental or systematic effects, is provided by the "*B*-mode" signal. This is an additional degree of freedom in the data, which is not produced by gravitational lensing, so is expected to be zero in the absence of systematics[13]. Assuming a Gaussian noise distribution, the relative number of pixels above and below the first contour in the *B*-mode suggests that this contour is equivalent to a ~2.9σ detection threshold including both statistical and systematic noise.

---

*[1]California Institute of Technology MC105-24, 1200 E. California Blvd., Pasadena, CA 91125, USA. [2]Jet Propulsion Laboratory, Pasadena, CA 91109, USA. [3]Laboratoire d'Astrophysique de Marseille, 13376 Marseille Cedex 12, France. [4]Max-Planck-Institut für extraterrestrische Physik, Giessenbachstraße, 85748 Garching, Germany. [5]Institute for Astronomy, Blackford Hill, Edinburgh EH9 3HJ, UK. [6]Service d'Astrophysique, CEA/Saclay, 91191 Gif-sur-Yvette, France. [7]Space Telescope Science Institute, 3700 San Martin Drive, Baltimore, MD 21218, USA. [8]Institut d'Astrophysique de Paris, Université Pierre et Marie Curie, 98 bis Boulevard Arago, 75014 Paris, France. [9]Physics Department, Ehime University, 2-5 Bunkyou, Matuyama, 790-8577, Japan. [10]Department of Physics and Astronomy, University of Waterloo, Waterloo, Ontario N2L 3G1, Canada.*

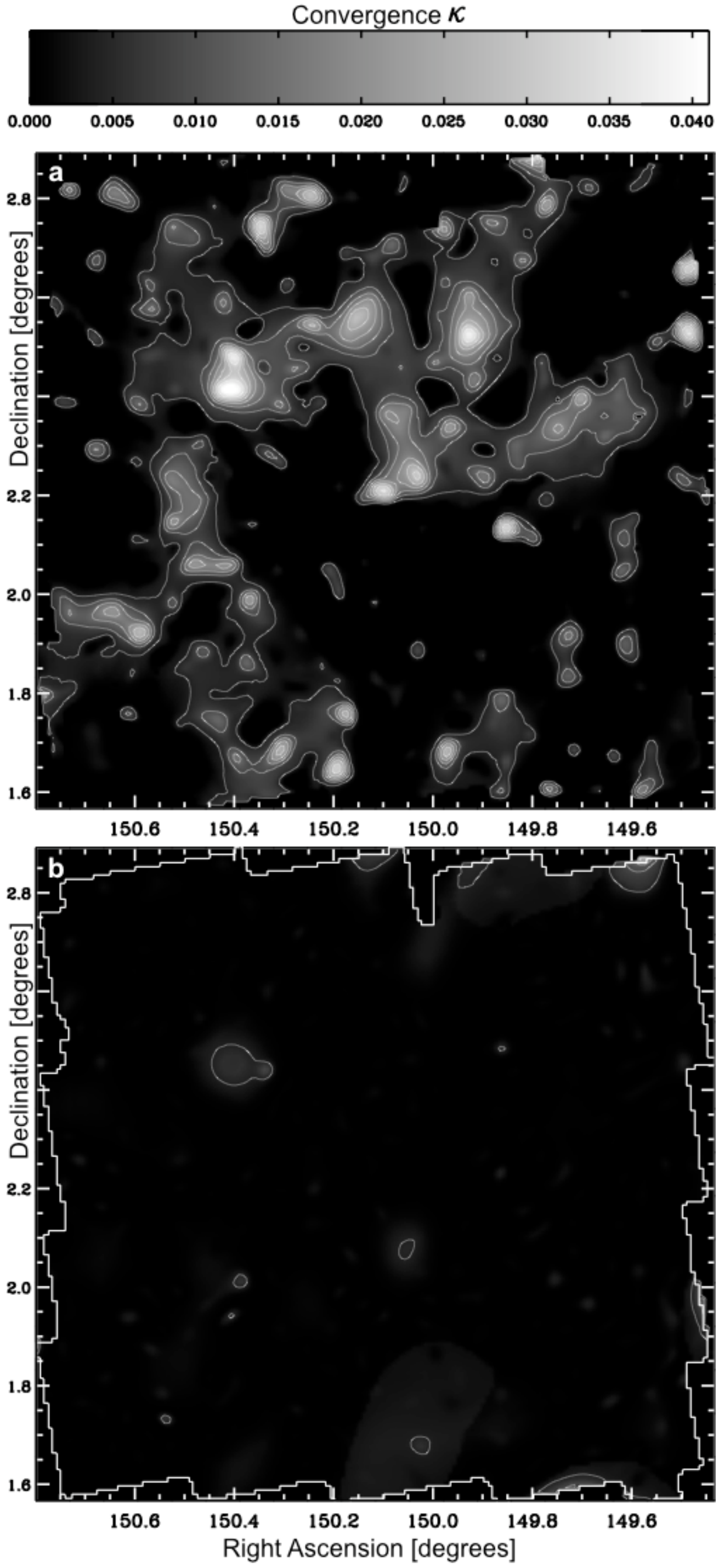

**Figure 1 | Map of the dark matter distribution in the 2 square degree COSMOS field. a**, the linear greyscale shows the "*E*-mode" lensing convergence field κ, which is proportional to the projected mass along the line of sight. Contours begin at 0.4% and are spaced by 0.5% in κ. **b**, the absolute value of the "*B*-mode" signal, shown with the same greyscale and contour levels, provides a realisation of the noise level in the map, plus contamination from uncorrected systematic effects; the bold outline traces the region observed with *HST*.

Gravitational lensing measurements have an unusual sensitivity to the distance of the influencing mass (Fig. 2), in contrasts to the familiar $\propto 1/d^2$ decline in luminosity of optically visible sources. Like an ordinary glass lens, a gravitational lens is most efficient when placed half way between the source and the observer – so lensing is sensitive to neither very distant *nor very nearby* structures. In order to compare the distribution of dark matter with that of baryons, we have mimicked this effect by appropriately weighting the foreground galaxies as a function of their redshift (cosmological distance). The number density of these independent galaxies, and the mass contained in their stars (estimated from their colour) provide two matched tracers of baryons. Deep *x*-rays observations with the *XMM-Newton* satellite additionally highlight concentrations of hot, dense gas.

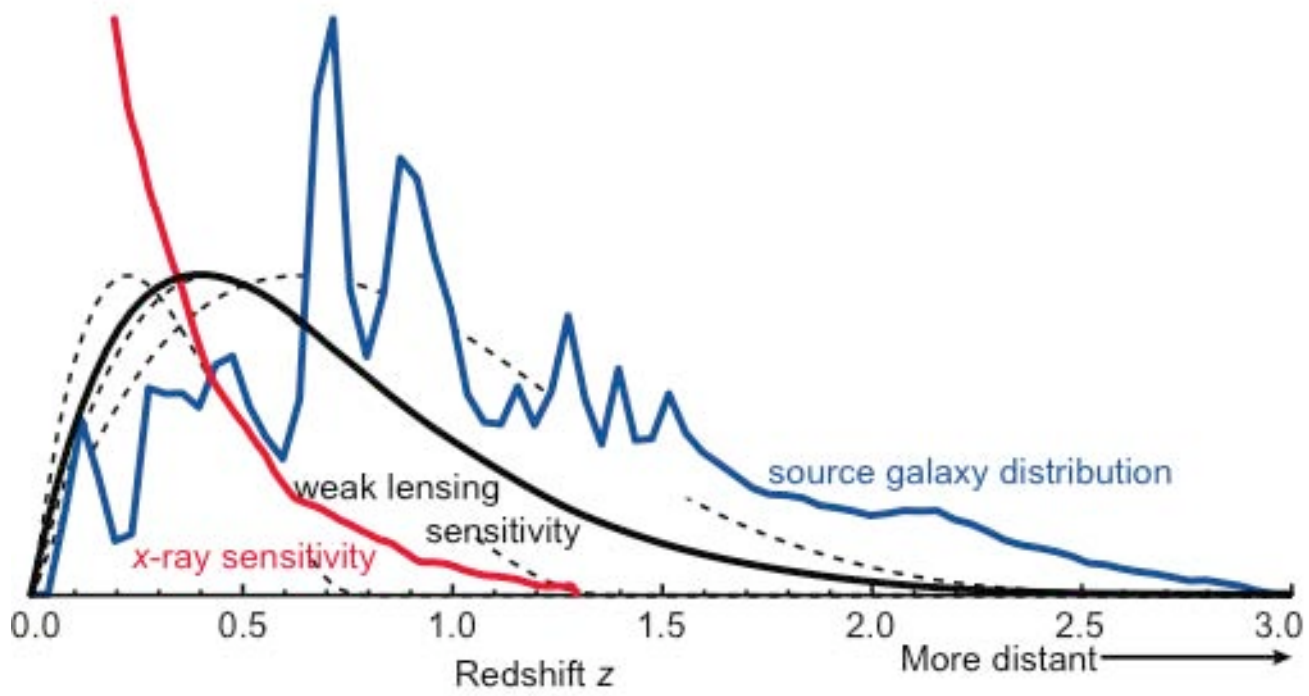

**Figure 2 | Sensitivity of probes of large-scale structure, as a function of distance.** The blue line shows the distribution of photometric redshifts for the source galaxies. The solid black line shows the sensitivity function of the lensing material for this source population (arbitrarily normalised to peak at unity) whereas the dashed lines show the equivalent sensitivities for the tomographic analysis. The red line shows the (arbitrarily normalised) sensitivity of the *x*-ray detections. Since the survey volume is a cone, the effective volumes peak at z=0.7 (lensing) and z=0.4 (*x*-rays).

The most prominent peak in the projected, 2D distributions of all four tracers (Fig. 3), is a single cluster of galaxies at (149.93, 2.52) and redshift $z$=0.73. *X*-rays are sensitive to the square of the electron density, so preferentially highlight the central cluster core. This cluster has an *x*-ray temperature $kT_X = 3.51^{+0.60}_{-0.46}$ *keV* and luminosity $L_X$=(1.56±0.04)×10$^{44}$ *erg/s* (0.1→2.4 *keV* band)[14]. If the cluster gas distribution were in hydrostatic equilibrium ("relaxed"), this would imply a mass of (1.6±0.4)×10$^{14}$ $M_{\rm sun}$ within an $r_{500}$ radius of 1.4 *arcminutes*. However, the cluster is clearly still growing[14]. Gravitational lensing is linearly sensitive to mass, and reveals an extended dark matter halo around this cluster, which in turn lies at the nexus of several filaments. The lensing mass of the full halo is (6±3)×10$^{15}$ $M_{\rm sun}$. It is possible that such a large value includes a contribution from additional mass directly in front of the cluster, at redshifts where lensing is more sensitive. Similar projection effects might also explain the twin lensing peaks without obvious baryonic counterparts near (150.3, 2.75). Weak lensing analysis is very sensitive, and the map could also have been perturbed by finite-field edge effects or isolated defects in our model of the telescope's Point Spread Function that are difficult to detect individually.

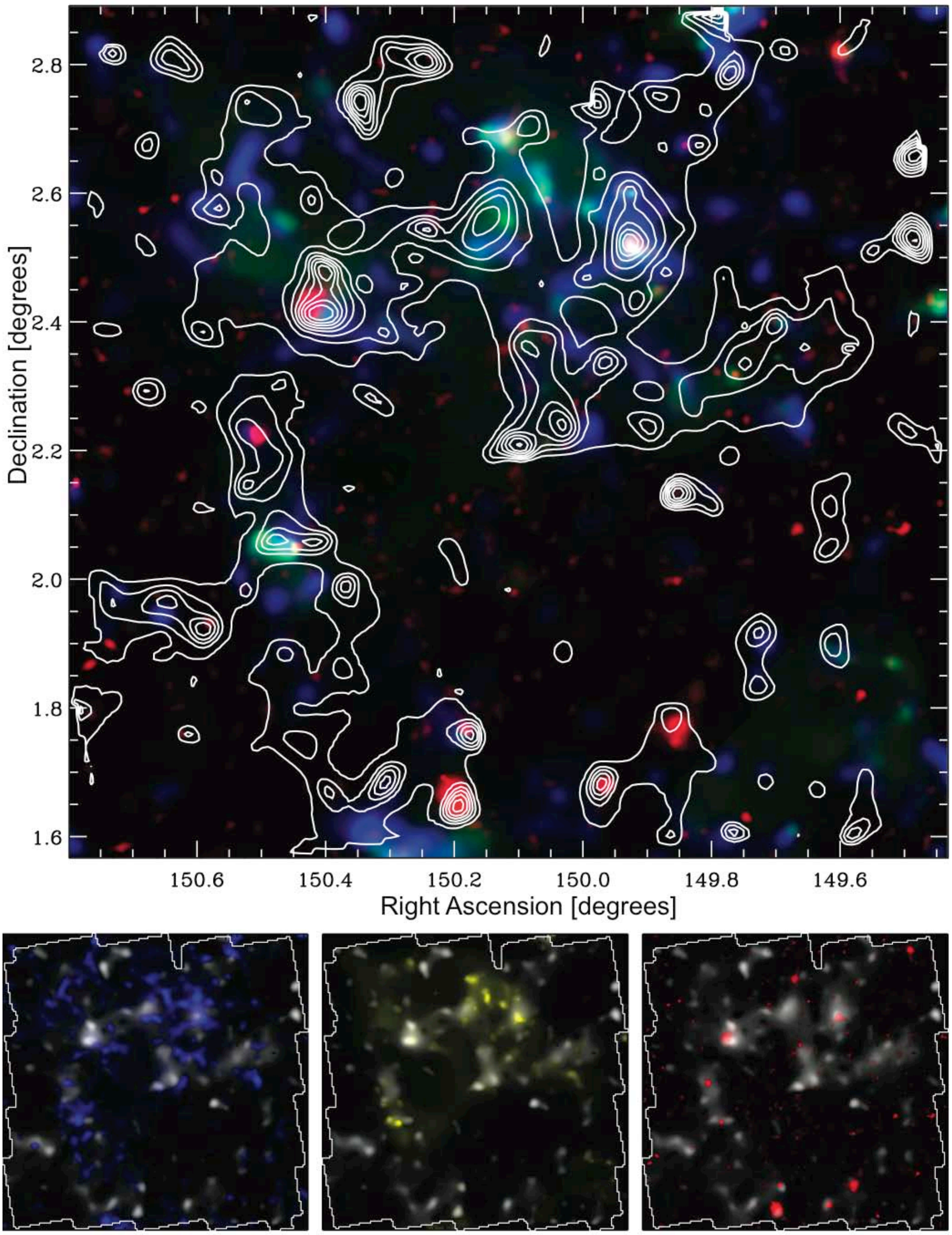

**Figure 3 | Comparison of baryonic and non-baryonic large-scale structure.** The total projected mass from weak lensing, dominated by dark matter, is shown as contours in panel **a** and as a linear grey scale in panels **b**, **c** and **d**. Independent baryonic tracers comprise (i) stellar mass (blue, colour scale peaks at $2.3\times10^{11}$ $M_{\rm sun}$ $deg^{-2}$ within $\Delta z$=0.1), (ii) galaxy number density (yellow, peak at $1.4\times10^{5}$ $deg^{-2}$ within $\Delta z$=0.1) seen in optical and near-IR light (adjusted to the redshift sensitivity function of the lensing mass), and (iii) hot gas (red, peak at $2.6\times10^{-14}$ $erg/s/cm^2/arcmin^2$) seen in *x*-rays after removal of point sources.

A statistical comparison across the entire map shows that baryons follow the distribution of dark matter even on large scales. The linear regression correlation coefficient $r$ of lensing mass with stellar mass is 0.42 and $r$ of lensing mass with galaxy number density is 0.47. The correlation with $x$-ray flux is somewhat lower, 0.30, consistent with the presence of filamentary structure outside the cluster cores. The map reveals overdense regions that are topologically connected but insufficiently dense to generate $x$-ray emission. These filaments are *not* a smoothing artefact, and cannot be reproduced by adding noise to the (square root of the) $x$-ray image then smoothing in a comparable way. We identify three distinct sets of environments, with a stark mass contrast. Filaments, defined as regions outside clusters but with lensing magnification $\kappa>0.4\%$, contain a 2.0× higher projected number density of galaxies compared to voids, and a 1.5× deficit compared to $x$-ray luminous clusters. Filaments have gravitationally collapsed along two axes, but clusters have continued to collapse along the third, and the latter ratio is expected to be ~5 for filaments in the plane of the sky[15]. The observed value is lowered by noise in the mass reconstruction, as well as partial alignment of filaments along our line of sight.

The 3D distribution of dark matter, and hence its time-dependent growth, can be visualised (Fig. 4) by splitting the background source galaxies into discrete redshift bins[16]. We have chosen bins so that the resulting foreground lensing sensitivities peak at redshifts $z \sim 0.3$, 0.5 and 0.7 (Fig. 2). These functions overlap slightly, so some structures can be faintly seen in successive slices. Catastrophic failures in photometric redshift measurement potentially mix slices further, although we have developed a method that minimises this effect in the lensing analysis. The massive $z$=0.73 cluster is indeed part of a much larger 3D structure, including a filament partially aligned with our line of sight, which will increase its 2D projected mass but not affect the $x$-ray flux. The corresponding $B$-mode maps (supplementary Fig. 4) suggest that the second contours here have roughly the same $3\sigma$ significance as the first contour in Figure 1. A full 3D reconstruction of the mass distribution (Fig. 5) is obtained from the differential growth of the lensing signal between many thin slices separated by $\Delta z = 0.05$[17,18]. The evolution of this distribution is driven by the battle between gravitational collapse and the accelerating expansion of the universe.

The independent probes of large-scale structure paint a remarkably consistent picture of the universe on large scales. The contracting filamentary network resembles predictions from $n$-body simulations of structure formation dominated by the gravitational collapse of cold dark matter from small density perturbations in the early universe[9]. By directly probing the distribution of mass, space-based weak lensing measurements offer the potential to directly link observations to theories that are concerned mainly with collisionless dark matter and gravity. Indeed, the resemblance of figure 5 to the first 3D maps of the large-scale distribution of baryonic matter made by the *Point Source redshift Catalogue*[19] (PSCz) fifteen years ago demonstrates profound progress in observational astronomy.

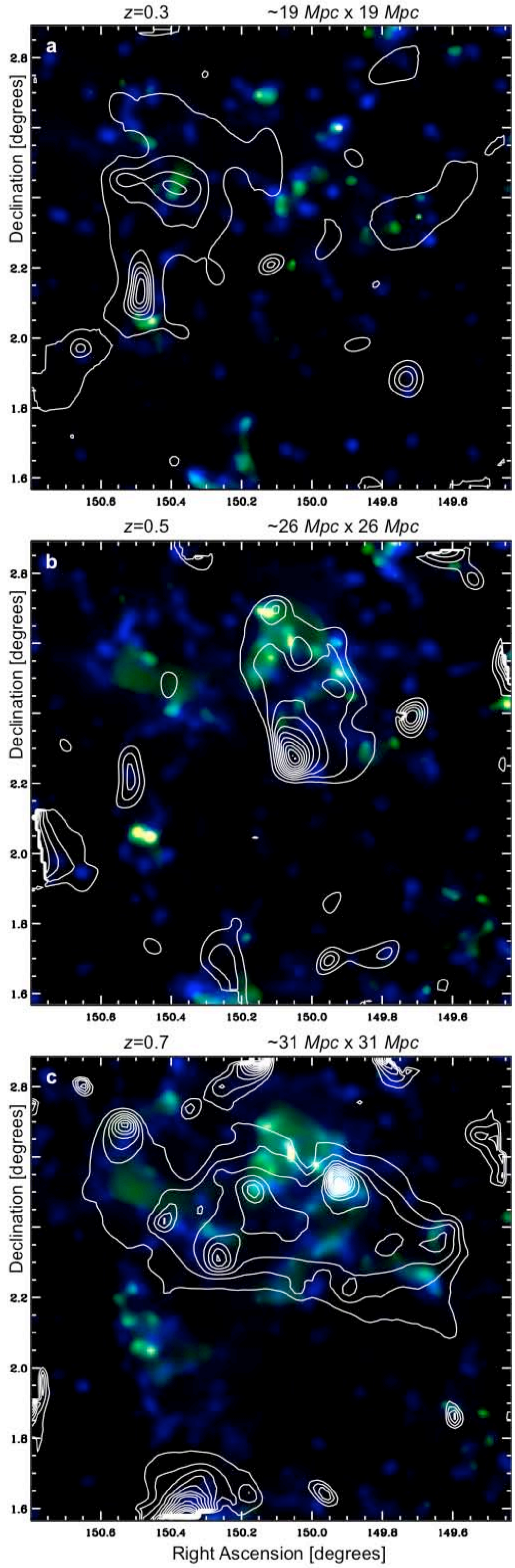

**Figure 4 | Growth of large-scale structure.** Slices through the evolving distribution of dark matter, created by splitting the background source galaxy population into discrete redshift slices. The sensitivity functions of the mass reconstruction peak at redshifts of ~0.3, ~0.5 and ~0.7 from panels **a** to **c**. Contours show the lensing convergence, in steps of 0.33%. A linear green colour ramp shows the distribution of galaxies, and blue their stellar mass, both weighted with matched redshift sensitivity functions.

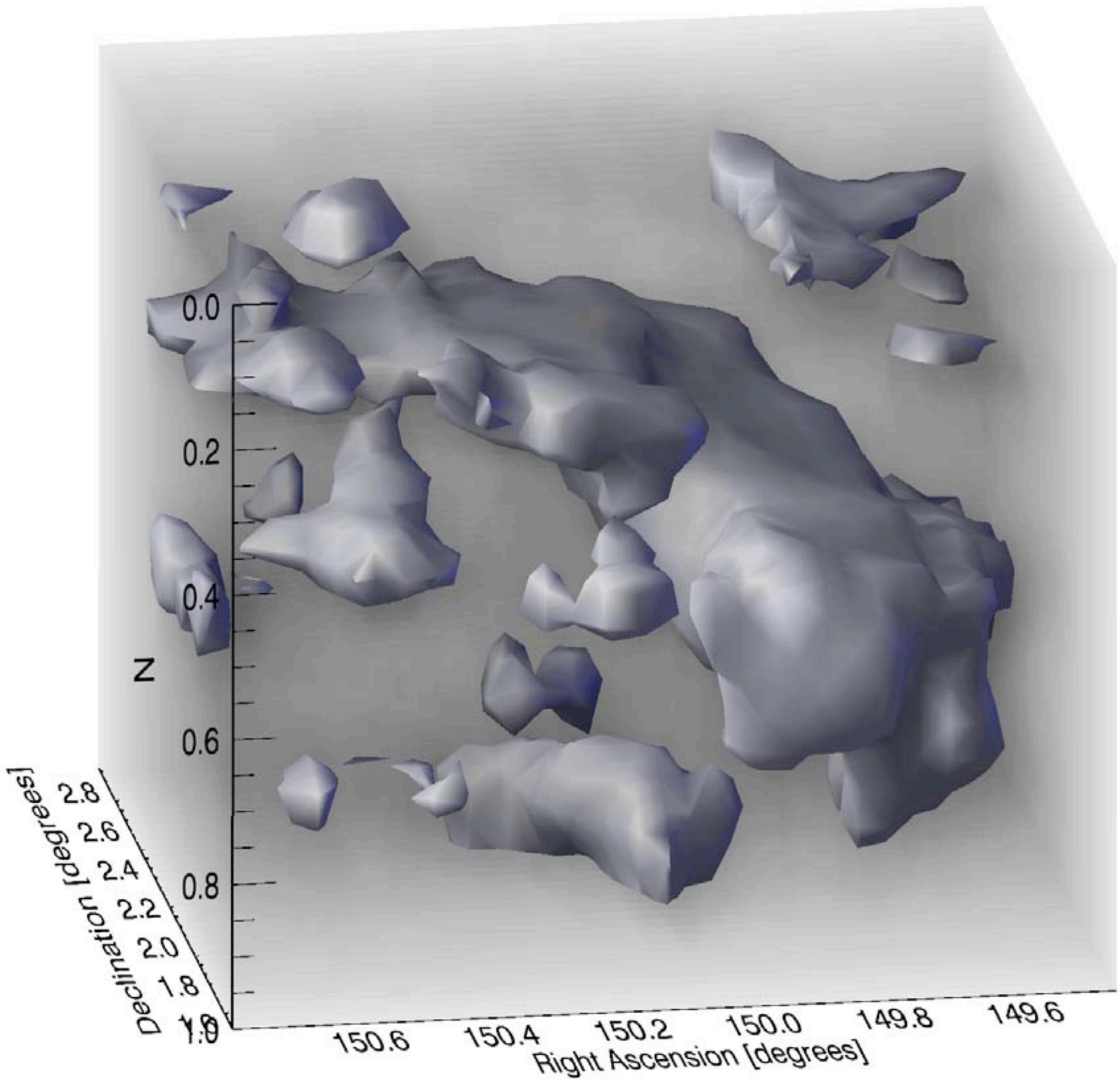

**Figure 5 | 3D reconstruction of the dark matter distribution.** The three axes correspond to Right Ascension, Declination, and redshift: with distance from the Earth increasing towards the bottom. The redshift scale is highly compressed, and the survey volume is really an elongated cone. An isodensity contour has been drawn at a level of $1.6\times10^{12}$ $M_{sun}$ within a circle of radius 700 *kpc* and $\Delta z$=0.05. This was chosen arbitrarily to highlight the filamentary structure. The faint background shows the full distribution, with the level of the grey scale corresponding to the local density. Additional views are provided in supplementary Fig. 7.

## METHODS

**Shear measurement.** The depth and exquisite resolution of *Hubble Space Telescope* images enable us to resolve the shapes of 71 galaxies per square arcminute in the *F814W* (approximately *I*-band) filter, with a median AB magnitude of 25.1. We use the RRG method[20] to deconvolve the galaxy shapes from the telescope's point spread function. Our processing pipeline has been calibrated on simulated *HST* images, and found to recover shear from galaxies in a wide range of size and flux, with less than 6% bias[12].

**Mass reconstruction.** The observed shear field is then converted into "convergence" (the cumulative magnification of all lenses along a line of sight), which is proportional to the two dimensional, projected mass[21]. The conversion is non-local, and finite-field effects introduce some defects near the edge of the map. Diversity in galaxies' intrinsic morphologies propagates into shot noise in the mass map, which we reduce via a multiscale filtering method, based on the *à trous* wavelet transform[22], and tuned to balance completeness with minimal false detection of spurious signals[23]. Our high surface density of resolved galaxies permits a mass reconstruction with an unprecedented minimum wavelet scale (maximum resolution) of 1.2 *arcminutes* FWHM in the projected map, and 2.4 *arcminutes* FWHM in the tomographic analysis. In practice, the achieved resolution of the wavelet reconstruction varies spatially with the local signal strength. Noise properties vary on the different wavelet scales, and are summed in a complex fashion that also depends upon the local signal. The noise is best quantified via the corresponding *B*-mode maps.

**Charge Transfer Efficiency correction**. One particularly troublesome systematic is introduced by image trailing during CCD detector readout, due to radiation damage that degrades their Charge Transfer Efficiency (CTE)[24]. The arrangement of the *ACS* CCDs produces a spurious sawtooth *E*-mode convergence pattern (with an average peak signal of ±0.3% and a pitch of 3.3', corresponding to the *ACS* field of view); but no corresponding *B*-mode (supplementary Fig. 2). We subtract a model[25] of the spurious shear signal from the galaxy catalogue, which incorporates parameters of galaxy flux, position on the CCD and date of exposure. After this correction, and the removal of high-frequency information by wavelet filtering, the spurious *E*-mode convergence signal is less than 0.1% throughout the map (supplementary Fig. 3). We note that the sharp sawtooth pattern is strongest on small scales, and systematic CTE deterioration thus limits the resolution of mass reconstruction from *HST*-based observations of weak lensing, at a level only just below the statistical limit set by the finite number density of resolved galaxies.

**Photometric redshift measurement.** Extensive follow-up observations of the COSMOS field with the *Subaru*, *CFHT*, *CTIO*-4*m* and *KPNO*-4*m* telescopes[26] has provided optical and near-IR imaging in 15 bands ranging from $u^*$ to $K_s$. Such multicolour data constitutes low-resolution spectroscopy, and we have used a Bayesian template-fitting method[27,28] to estimate the redshift and stellar mass of each galaxy (supplementary Fig. 5). The depth of the follow-up observations ensures completeness in stellar mass down to $7\times10^9$ $M_{\rm sun}$ at $z<1.05$[11]. In the source galaxy redshift distribution (Fig. 2), the foreground peaks below $z$=1.2 all correspond to known structures in the COSMOS field.

For galaxies fainter than those with an AB magnitude of 24.5 in the *F814W* filter, a characteristic degeneracy exists between galaxies at $0.1<z<0.3$ and $1.5<z<3.2$ without real spectroscopy, due to confusion between the 4000Å break and coronal absorption features[26]. For the purposes of weak lensing, this degeneracy is not symmetric. Distant galaxies are viewed after significant distortion and, if placed erroneously at low redshift, would create spurious power in the nearby universe that echoes more distant structures. On the other hand, very nearby galaxies are almost unlensed so, if placed incorrectly at high redshift, they merely dilute the signal. To deal with this redshift degeneracy, we study the joint redshift probability distribution function for each galaxy. For those with best-fitting redshifts $z<0.4$, if *any* probability exists above $z>1.5$, we move the galaxy to the weighted mean of the redshift probability integrated above $z=1.5$. This places all uncertain galaxies in the same place, and contracts two problems into the less troublesome one. We then statistically estimate the overpopulation of high redshift slices by comparing their apparent density of galaxies to that expected from the known galaxy luminosity function[28,29] (supplementary Fig. 6).

**Supplementary Information** accompanies the paper on www.nature.com/nature. More details about the COSMOS survey are available at www.astro.caltech.edu/~cosmos.

**Acknowledgements** Based on observations with the NASA/ESA Hubble Space Telescope, obtained at the Space Telescope Science Institute, which is operated by the Association of Universities for Research in Astronomy, Inc. (AURA). Also based on data collected from: the XMM-Newton, an ESA science mission with instruments and contributions directly funded by ESA member states and NASA; the Subaru Telescope, which is operated by the National Astronomical Observatory of Japan; the European Southern Observatory, Chile; Kitt Peak National Observatory, Cerro Tololo Inter-American Observatory, and the National Optical Astronomy Observatory, which are operated by AURA under cooperative agreement with the American National Science Foundation; the National Radio Astronomy Observatory, which is a facility of the American National Science Foundation operated under cooperative agreement by Associated Universities, Inc.; and the Canada-France-Hawaii Telescope operated by the National Research Council of Canada, the Centre National de la Recherche Scientifique de France and the University of Hawaii. We gratefully acknowledge the contributions of the entire COSMOS collaboration consisting of more than 70 scientists. We thank T. Roman, D. Taylor, and D. Soderblom for help scheduling the extensive COSMOS observations; and A. Laity, A. Alexov, B. Berriman and J. Good for managing online archives and servers for the COSMOS datasets at NASA IPAC/IRSA. This work was supported by grants from NASA (to NS and RM).

**Author Contributions** AK processed the raw HST data, and JPK masked defects in the image. AL, JR and RM catalogued the positions and shapes of galaxies. YT and SS obtained multicolour follow-up data, which was processed and calibrated by SS, PC, HM and HA. PC determined galaxies' redshifts, and BM their stellar mass. NS constructed maps of stellar mass and galaxy density. AF processed the $x$-ray image and removed point sources. RM and AR produced the 2D and tomographic mass maps; JLS and SP developed the wavelet filtering technique. DB and AT produced the 3D mass reconstruction. JT, AF, RE and RM compared the various tracers of large-scale structure.

**Author Information** Correspondence and requests for materials should be addressed to RM (rjm@astro.caltech.edu).

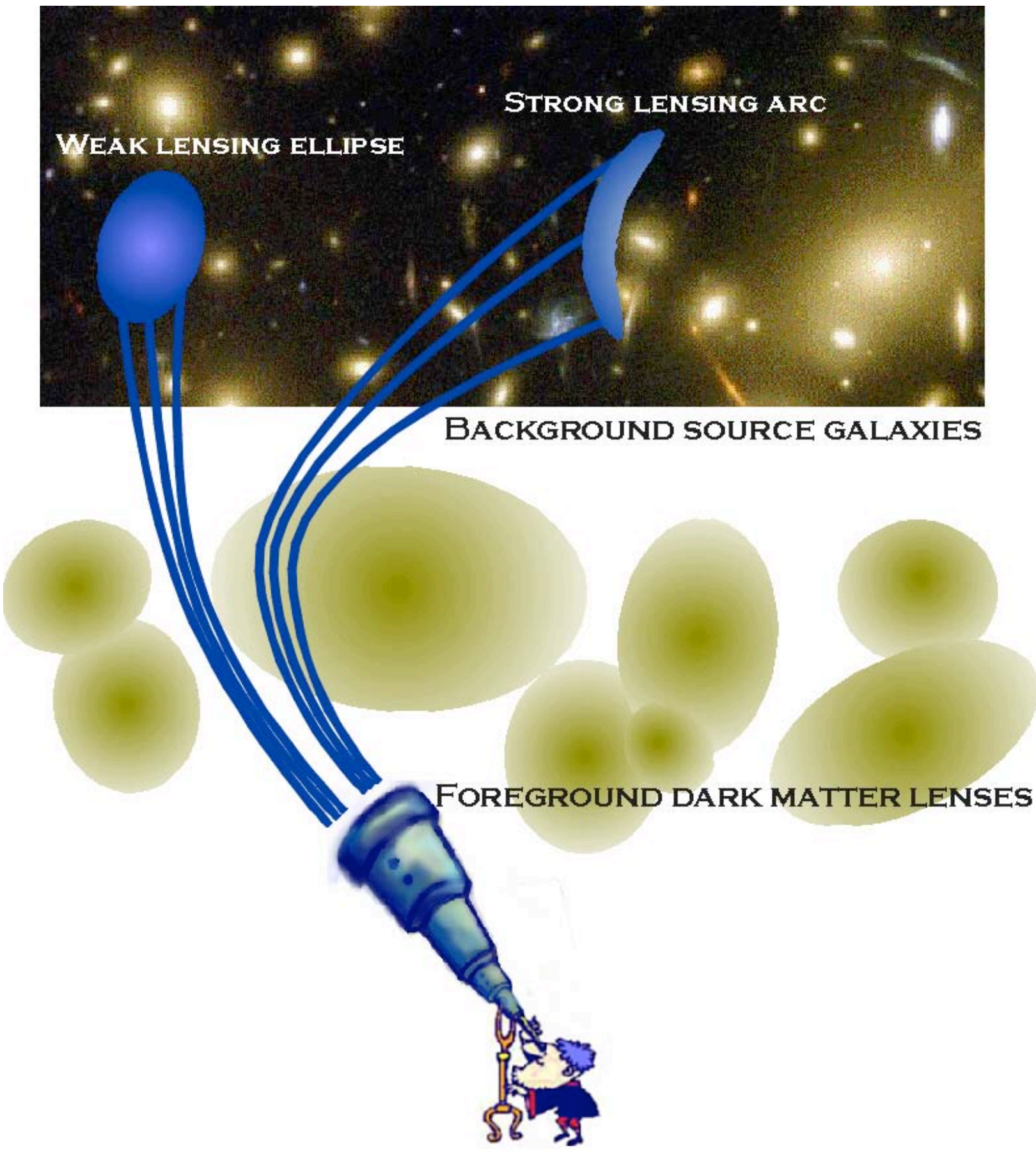

**Figure S1 | Cartoon illustrating the process of gravitational lensing.** Gravitational lensing by foreground mass distorts the shapes of distant galaxies, regardless of the nature of that mass. It is thus sensitive to the dominant contribution of dark matter. Measuring these distortions to infer the intervening distribution of dark matter is like studying bent lines of text through a magnifying glass to study the properties of the glass lens. Along lines of sight including the most dense mass concentrations, *strong gravitational lensing* distorts individual background galaxy shapes into multiply-imaged giant arcs. Along a more typical line of sight, the modest *weak gravitational lensing* induces a 2-3% change of ellipticity. In this regime, the lensing distortion is smaller than the typical variation between galaxy morphologies, and it cannot be measured from a single galaxy because its intrinsic shape is not observable. However, the signal can still be detected statistically. The intrinsic shapes of adjacent galaxies are unrelated so, in the absence of lensing, they should have no preferred orientation. On the other hand, when adjacent galaxies have been lensed by approximately the same intervening mass, the weak lensing signal can be recovered from the degree of correlation between their shapes.

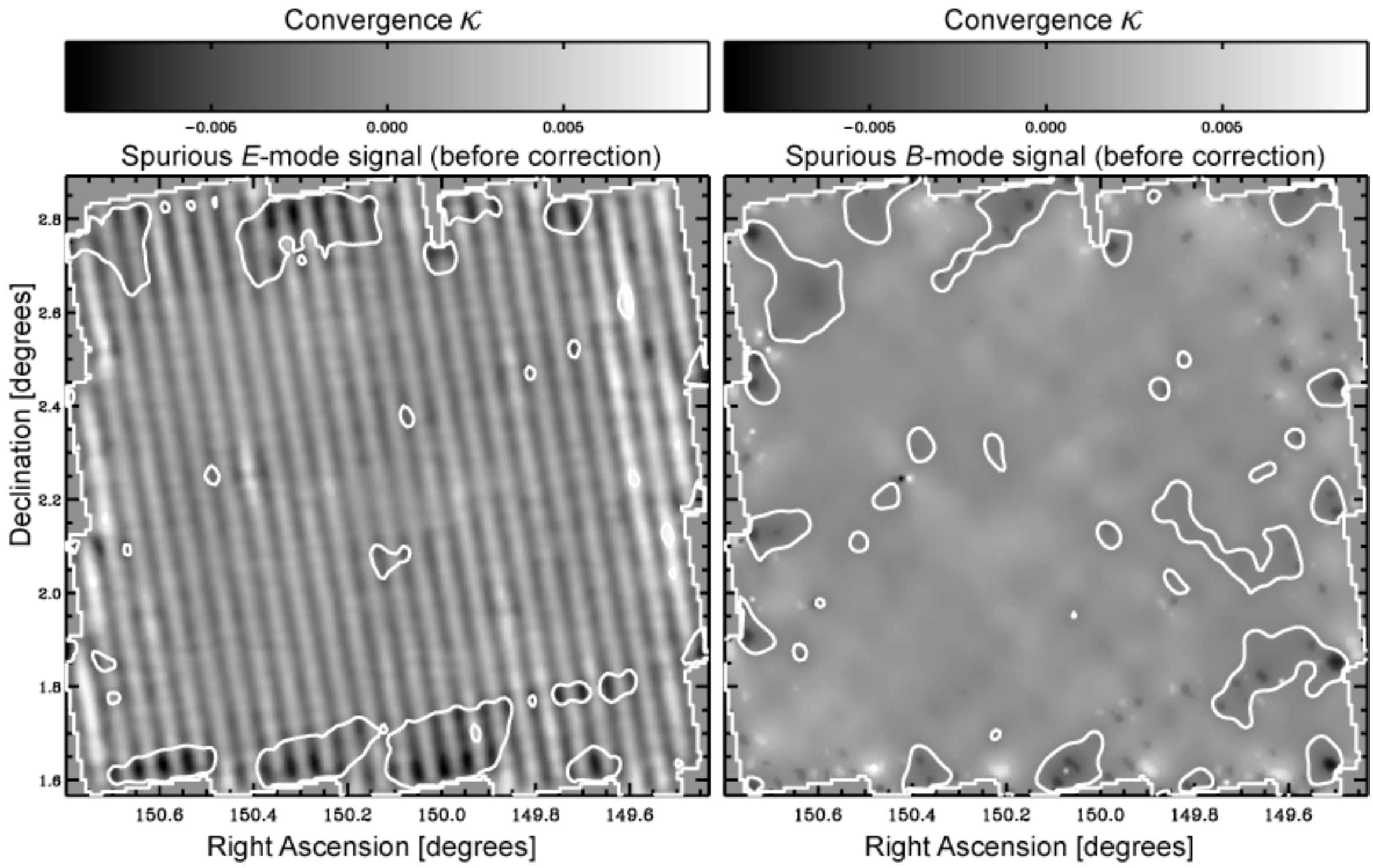

**Figure S2 | Spurious signal due to imperfect Charge Transfer Efficiency (CTE) *before correction*.** The continual bombardment of spacecraft electronics by high energy particles degrades detectors' CTE. During CCD readout, defects in the silicon substrate act as charge traps that trail the image in the readout direction, mimicking a weak lensing signal. Deterioration continued during the two years of observations, and the spurious signal is worse at the edges of the field than in the middle, because of the approximately spiral pattern in which the data were obtained. CTE trailing is of particular concern in this data set because the arrangement of the *HST ACS* CCDs places spurious signal entirely into the measured *E*-mode (left panel), with no coherent *B*-mode counterpart (right panel). In this respect, it is unlike most other expected sources of observational systematics. To highlight the oscillatory, sawtooth pattern created by the CCD configuration, the greyscale image shows a reconstruction of our model of the spurious signal with twice the resolution (0.6' FWHM) to that used in the mass maps (Fig. 1 in the main article). This most clearly reveals the sawtooth pattern, but is only possible because our CTE model contains no noise. The contours, which are drawn at the same levels as those in Fig. 1, show the reconstruction after the same amount of smoothing used the real mass map. After this smoothing, the mean peak height across the reconstruction is 0.1%. The conversion of shear to convergence involves a complex iteration over scales, and this provides the most appropriate estimate of the (uncorrected) signal due to CTE effects.

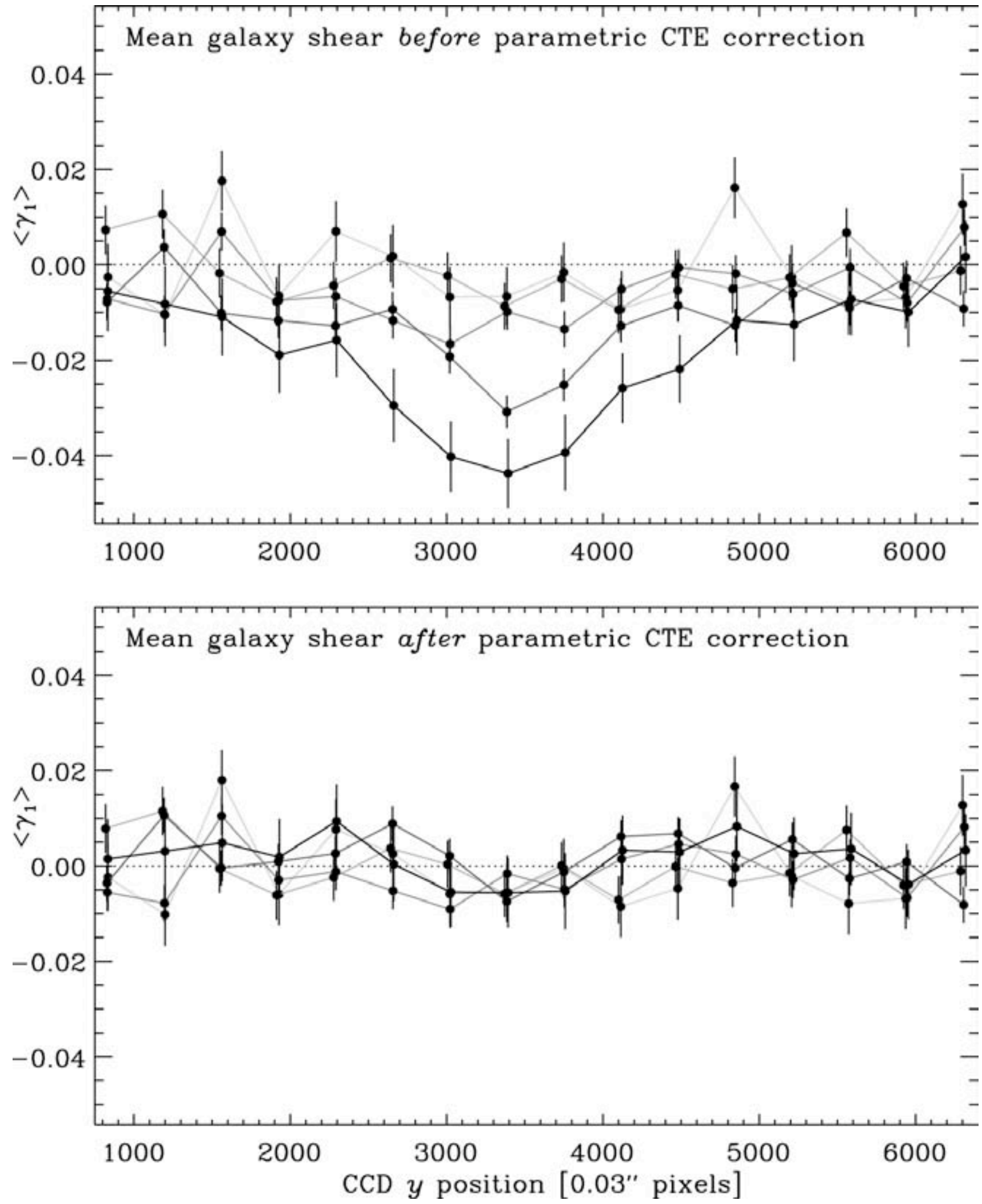

**Figure S3 | Correction for Charge Transfer Efficiency (CTE).** The spurious CTE signal can be measured independently by directly stacking galaxies from many different pointings as a function of their position on the CCD. Since the images cover different patches of sky, the cosmological signal averages away. The top panel shows the spurious vertical (negative) elongation of galaxies before correction. The ACS CCDs have readout registers at the top and bottom of the field of view, so the CTE signal is worst in a band across the middle. The top (light coloured) line is for bright galaxies between *F814W* magnitudes 22 and 23; the other lines are in increments of one magnitude, continuing to the faintest and most affected galaxies between magnitudes 26 and 27 in the bottom (dark coloured) line (s.e.m. errors). The bottom panel shows the residual signal after applying our parametric correction to the shear catalogue. This reduces the signal by a further factor of at least 5 (stronger statements would require an even larger survey to test). Using this model correction in the mass reconstruction takes the peak convergence level below 0.1%. Reproduced from Rhodes et al. (2007), with permission from the Astrophysical Journal.

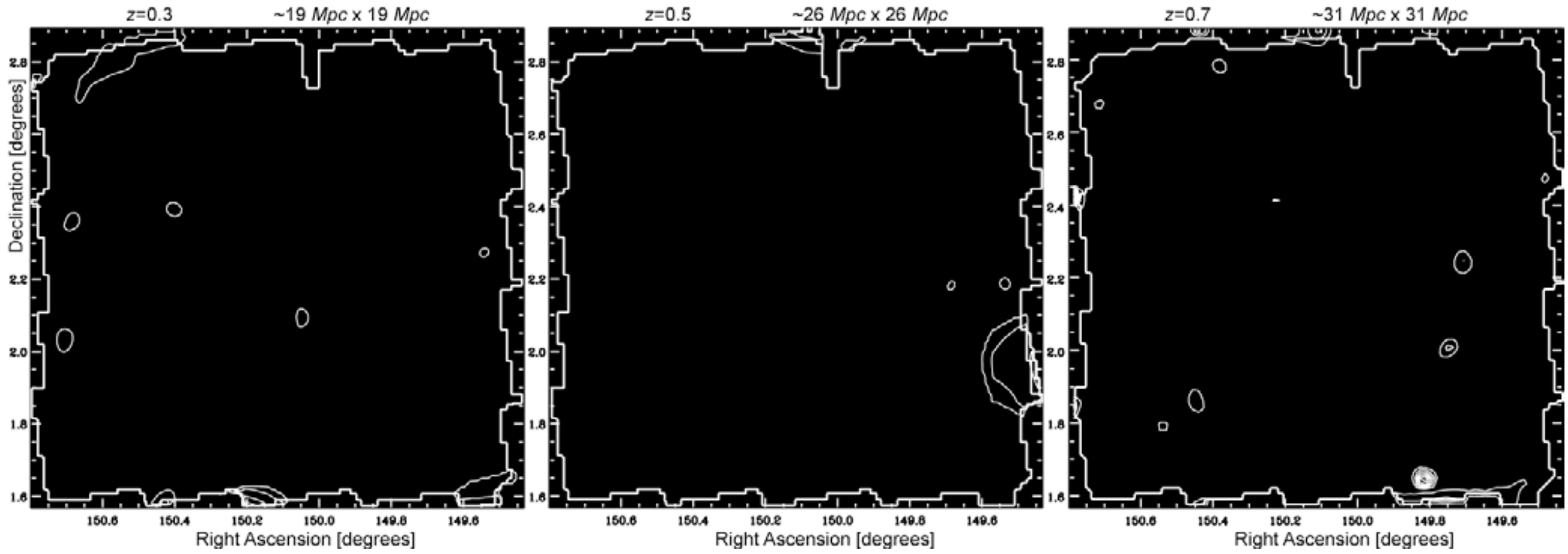

**Figure S4 | Realisation of noise level in the tomographic mass reconstructions.** These show the "*B*-mode" signal from galaxy populations restricted to redshift slices, and are expected to be consistent with zero in the absence of systematic effects. Contours are drawn at the same levels as those in the "*E*-mode" weak lensing maps (Fig. 4). Judging by the ratio of the pixels above and below the contours as before, and assuming Gaussian noise distribution, the second contour corresponds to ~3σ significance.

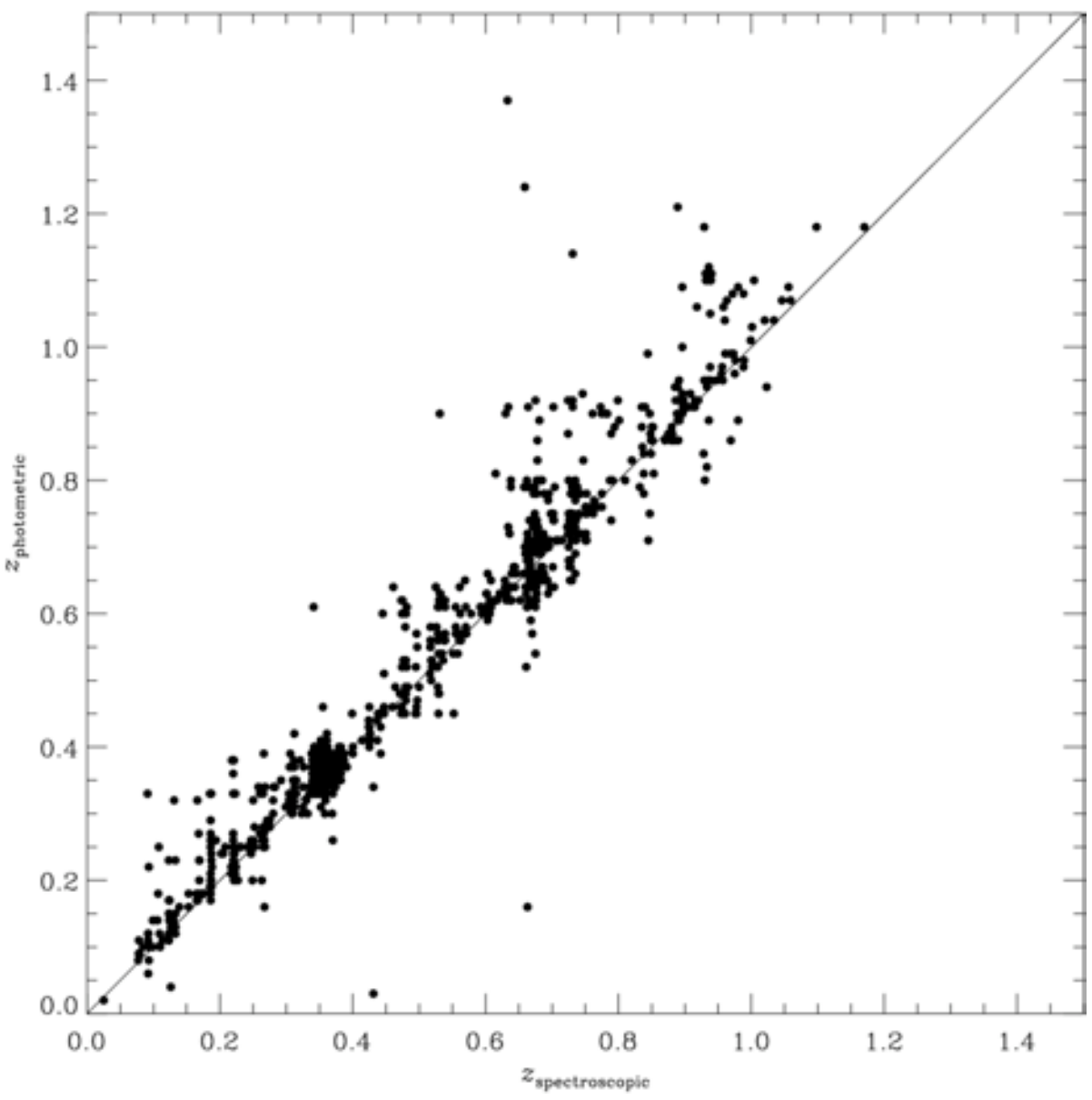

**Figure S5 | Photometric redshift accuracy for bright galaxies.** This demonstrates the $\delta z \approx 0.15$ performance of our photometric redshift estimation in a sample of 822 relatively bright (*F814W*>22.5) galaxies for which spectroscopic redshift are available. Accuracy inevitably degrades for fainter galaxies, which make up the majority of our source population. In particular, these are subject to "catastrophic" failures, which we treat asymmetrically using our novel statistical method described in the main article.

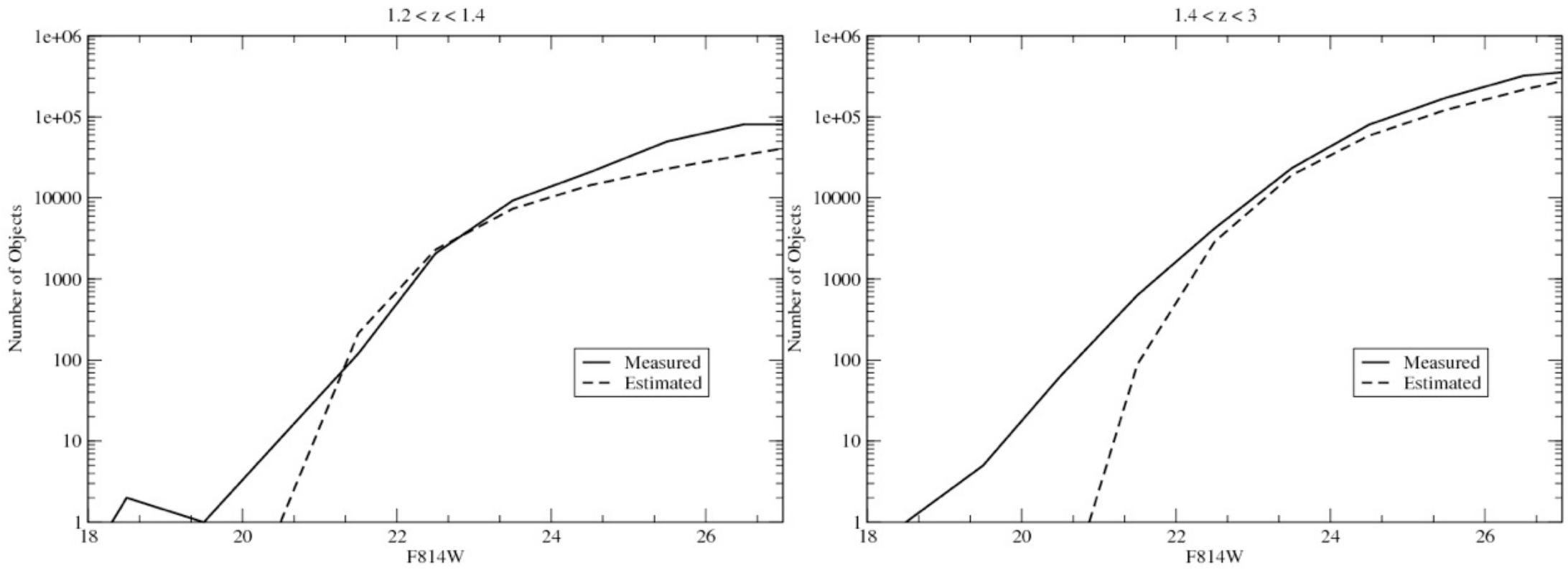

**Figure S6 | Measured and expected number counts of faint galaxies.** Shown as a function of apparent magnitude for galaxies in redshift bins 1.2<*z*<1.4 and 1.4<*z*<3. Notice the two are in good agreement except at the bright and faint ends. This disagreement is quantified in supplementary table S1.

| ***F814W* magnitude** | **Redshift** | | |
|---|---|---|---|
| | **1.0→1.2** | **1.2→1.4** | **1.4→3.0** |
| 20.5 | 2.8 | 10.6 | 926 |
| 21.5 | 1.2 | 1 | 7.1 |
| 22.5 | 1.05 | 1 | 1.46 |
| 23.5 | 1 | 1.26 | 1.22 |
| 24.5 | 1 | 1.45 | 1.36 |
| 25.5 | 1.5 | 2.17 | 1.41 |
| 26.5 | 1.5 | 2.4 | 1.48 |
| Total | 1.19 | 1.74 | 1.39 |

**Table S1 | Dilution of the lensing signal by the spurious inclusion of low redshift galaxies at high redshift.** The numbers are the ratio of the number of galaxies observed in each redshift bin (including those with incorrect photometric redshifts) to the number expected from external measurements of the galaxy luminosity function between *z*=1.2 and *z*=3. The total shows the dilution in each redshift bin, accounting for the distribution of galaxies of various magnitudes.

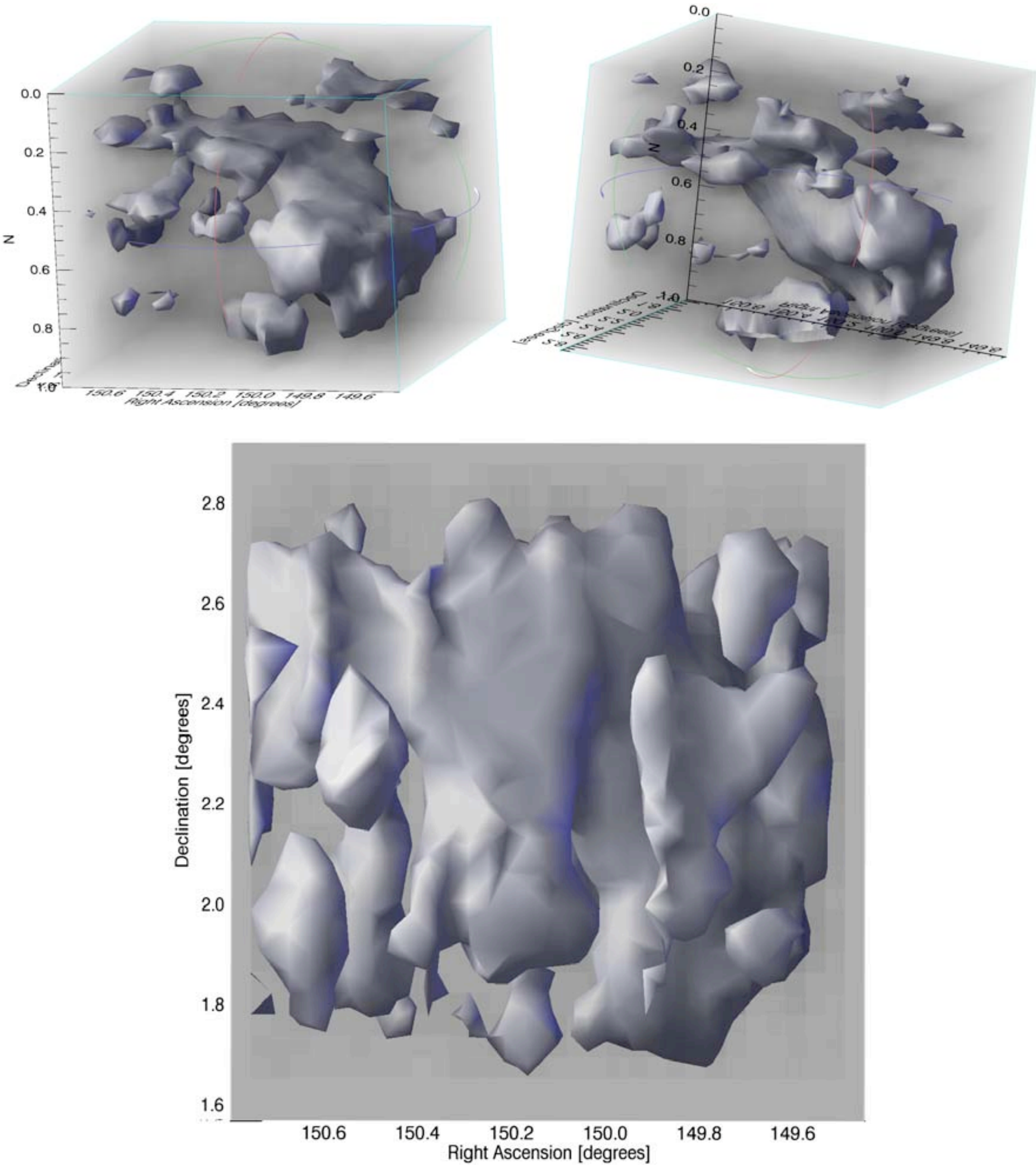

**Figure S7 | Additional views of the 3D mass reconstruction.** In the lower panel, it is seen from the perspective of Earth, as with the projected and tomographic mass maps in the main article.